\documentclass{article}
\newcommand{\be}{\begin{equation}}
\newcommand{\ee}{\end{equation}}
\newcommand{\de}{{\rm d}}
\newcommand{\ie}{{\rm i}}
\newcommand{\ex}[1]{{\rm e}^{#1}}
\newcommand{\sign}{{\rm sign}}
\newcommand{\tbl}[1]{\caption{#1}}
\newcommand{\toprule}{\hline}
\newcommand{\colrule}{\hline}
\newcommand{\botrule}{\hline}
\begin{document}

\title{Inhomogeneous Fixed Point Ensembles Revisited}

\author{Franz J. Wegner\\
Institute for Theoretical Physics, Ruprecht-Karls-University,\\
Philosophenweg 19, D-69120 Heidelberg, Germany}

\maketitle

\begin{abstract}
The density of states of disordered systems in the Wigner-Dyson classes
approaches some finite non-zero value at the mobility edge, whereas the density
of states in systems of the chiral and Bogolubov-de Gennes classes shows a
divergent or vanishing behavior in the band centre. Such types of behavior were
classified as homogeneous and inhomogeneous fixed point ensembles within a
real-space renormalization group approach. For the latter ensembles the scaling
law $\mu=d\nu-1$ was derived for the power laws of the density of states
$\rho\propto|E|^\mu$ and of the localization length $\xi\propto|E|^{-\nu}$.
This prediction from 1976 is checked against explicit results obtained
meanwhile.
\end{abstract}

%\body

\section{Introduction}

Some time ago I used real-space renormalization group arguments in analogy to
the cell model of Kadanoff\cite{Kadanoff} in order to investigate the critical
behavior\cite{Wegner76} close to the mobility edge of the Anderson
model\cite{Anderson}.
Two types of ensembles were considered, a homogeneous and an inhomogeneous one.

{\it Homogeneous fixed point ensemble (HFPE)}. This ensemble is homogeneous in
energy $\epsilon$. It is invariant under the transformation
$\epsilon\rightarrow\epsilon+{\rm constans}$. Since the density of states
$\rho$ stays constant during the renormalization group procedure the scale
change
\be
r\rightarrow r/b \quad{\rm implies} \quad \epsilon\rightarrow\epsilon b^d
\ee
with dimension $d$ of the system. We assume one relevant perturbation to this
system which grows like
\be
\tau\rightarrow\tau b^y.
\ee
Depending on the sign of $\tau$ the perturbation produces localized and extended
states, resp. This perturbation is added to the HFPF in a strength increasing
linearly in energy $E$
\be
\tau=cE,
\ee
where the mobility edge is taken at $E=0$, and extended states at $\tau>0$ and
localized ones for $\tau<0$. $c$ transforms under RG.

{\it Inhomogeneous fixed point ensemble (IHFPE)}. In this ensemble the scale
factors for length and energies are independent from each other. The ensemble
is inhomogeneous in the energy,
\be
r \rightarrow r/b, \quad \epsilon\rightarrow\epsilon b^y.
\ee
It is assumed that there is no relevant perturbation to such an ensemble.

Both ensembles yield power and homogeneity laws. The density of states obeys
\be
\rho_{\rm hom} = {\rm const}., \quad \rho_{\rm inh}\propto |E|^{\mu}, \quad
\mu = d/y-1. \label{expmu}
\ee
The localization length yields in both cases
\be
\xi \propto |E-E_c|^{-\nu}, \quad \nu=1/y. \label{expnu}
\ee
The low-temperature a.c. conductivity obeys the homogeneity relation
\be
\sigma(\omega,\tau) = \left\{\begin{array}{cl}
b^{2-d}\sigma(\omega b^d,\tau b^y) & \mbox{ HFPE} \\
b^{2-d}\sigma(\omega b^y,\tau b^y) & \mbox{ IHFPE}
\end{array} \right. .
\ee
One deduces the d.c. conductivity in the region of extended states
\be
\sigma(0,\tau) \sim \tau^s, \quad s=(d-2)/y = (d-2)\nu. \label{scaling_s}
\ee
What comes out correctly on the basis of these ideas? Not only the scaling and
homogeneity laws shown above can be deduced, but also such laws for averaged
correlations, including the inverse participation ratio and long-range
correlations between states energetically close to each other including those
in the vicinity of the mobility edge. What has to be added are averages of
matrix elements and of their powers for the transformation step by the linear
scale factor $b$ of the cell model.

A short historical digression may be allowed.
The oldest paper on the mobility edge i.e. the separation of localized and
extended states of a disordered system was given by Phil
Anderson\cite{Anderson}
(1958) (Well aware of possible complications by the Coulomb interaction he
considered the transition from spin diffusion to localized spin excitations).
It is a nice accident that its page number 1492 coincides with the year of
another important discovery.
Earlier papers on disordered systems, which became important for the
development of this field was Wigner's\cite{Wigner51,Wigner55,Wigner58}
Gaussian matrix ensemble (1951) for nuclei.
Probably the oldest paper on chiral systems is Dyson's paper\cite{Dyson53} on
disordered chains (1953). Other early contributions on disordered chains were
by Schmidt\cite{Schmidt} and arguments that states in one dimension are localized.\cite{Borland,Furstenberg}.
In 1962 Dyson gave the threefold classification of ensembles of orthogonal,
unitary and symplectic symmetry depending on the behaviour under time-reversal
invariance.\cite{Dyson62a,Dyson62b}.

Since these early developments a lot of progress has been made.
There are numerous calculations
for the behaviour around the mobility edge both analytic and numerical. I refer
to the review by Evers and Mirlin\cite{Evers}. 1979 marked important
break-throughs: 
The scaling theories of localization by Abrahams et al.\cite{Abrahams} and by
Oppermann and Wegner\cite{Oppermann} appeared.
The mapping onto a non-linear sigma-model was conjectured\cite{Wegner79},
brought into its bosonic-replica\cite{Schaefer80}, its
fermionic-replica\cite{Efetov80} and finally in
its supersymmetric\cite{Efetov83} form. A self-consistent approximation for the
Anderson transition was put forward by G\"otze\cite{Goetze}, Vollhardt and
W\"olfle\cite{Vollhardt80,Vollhardt82}.
A numerical renormalization scheme was devised by MacKinnon and
Kramer\cite{MacKinnon}.

Since then many more results and techniques were developped.
Here I mention only a few: The complete classification of ten symmetry classes
of random matrix theories, $\sigma$-models, and Cartan's symmetric
spaces was given by Zirnbauer and Altland\cite{Zirnbauer96,Altland}
and by Schnyder, Ryu, Furusaki, and Ludwig\cite{Schnyder08} after several
occurences of chiral and Bogolubov-de Gennes
classes\cite{Hikami,Oppermann87,Oppermann90,Wegner89,Gade91,Gade93,Slevin93,Verbaarschot93}.
These classes are listed in table \ref{tb1} since I will refer later to this
nomenclature.

\begin{table}[h]
\tbl{Symmetry classes of single particle Hamiltonians defined in terms of
presence or absence of time-reversal symmetry (TRS) and particle-hole symmetry
(PHS). Absence is denoted by 0, presence by the symmetry square $\pm 1$.
SLS indicates absence (0) and presence (1) of sublattice or chiral symmetry.
After ref.\cite{Schnyder08}}
{\begin{tabular}{cccccc} \toprule
System & Cartan & symmetry & TRS & PHS & SLS \\
 & nomenclature & & & & \\ \colrule
standard & A & unitary & 0 & 0 & 0 \\
(Wigner-Dyson) & AI & orthogonal & $+1$ & 0 & 0 \\
 & AII & symplectic & $-1$ & 0 & 0 \\ \colrule
chiral & AIII & unitary & 0 & 0 & 1 \\
(sublattice) & BDI & orthogonal & $+1$ & $+1$ & 1 \\
 & CII & symplectic & $-1$ & $-1$ & 1 \\ \colrule
Bogolubov- & D & & 0 & $+1$ & 0 \\
de Gennes & C & & 0 & $-1$ & 0 \\
 & DIII & & $-1$ & $+1$ & 1 \\
 & CI & & $+1$ & $-1$ & 1 \\
\botrule
\end{tabular}}
\label{tb1}
\end{table}

Transfer matrix approaches originally used for linear chains were developped for
the non-linear $\sigma$-model\cite{Efetov83} as well as for the for the
distribution function of the transfer matrix of chains with many channels
(DMPK-equation\cite{Dorokhov,Mello}). These techniques allowed the
determination of correlations, wave-function statistics and transport
properties. Such chains can have broad distributions of conductivities and even
cases of perfect transmissions were found\cite{Zirnbauer92,Mirlin94}.

In two dimensions some of these symmetry classes allow the inclusion of a
topological $\theta$-term. As observed by Pruisken et al.\cite{Levine,Pruisken}
the Wigner-Dyson unitary class with this term describes the integer quantum
Hall effect. Another term which may be added is a Wess-Zumino term. Such terms
are of importance in the study of disordered Dirac fermions, which appear in
dirty d-wave superconductors\cite{Nersesyan,Bocquet00,Altland02} and in
disordered 
graphene.\cite{Aleiner06,Khvesh06,McCann,Ostrovsky06,Ostrovsky07a,Ostrovsky07b}

Network models originally introduced by Shapiro\cite{Shapiro} are very useful
for the description of quantum Hall systems as first shown for the integer
quantum Hall effect in the Chalker-Coddington-model\cite{Chalker88}.

Obviously the HFPE applies to the Wigner-Dyson classes, whereas the
IHFPE applies to the chiral and the Bogoliubov-de Gennes classes.

The main object of this paper is the comparison of the scaling law for the IHFPE
\be
\mu=d\nu -1. \label{scalinglaw}
\ee
derived from (\ref{expmu}, \ref{expnu}) with various results meanwhile obtained.
I will shortly comment on the scaling law (\ref{scaling_s}) for the conductivity
in subsection \ref{cond}.

\section{One dimensional chains}

\subsection{Thouless relation}

Thouless\cite{Thouless72} following Herbert and Jones\cite{Herbert} considered a
one-dimensional chain governed by the Hamiltonian
\be
H=\sum_{i=1}^N \epsilon_i|i\rangle\langle i|
- \sum_{i=1}^{N-1} (V_{i,i+1}|i\rangle \langle i+1| +
V_{i+1,i}|i+1\rangle\langle i| ) \label{Hchain}
\ee
and found in the limit $N\rightarrow\infty$ that the function
\be
K(z) = \int\de x \rho(x) \ln(x-z) - \overline{\ln |V|}, \quad
-\pi<\arg\ln(x-z)<\pi
\ee
connects both the integrated density of states $I(E)$ and the exponential
decrease of eigenfunctions $\lambda(E)$ (inverse correlation length $\xi$)
\be
K(E\pm\ie 0) =\lambda(E) \mp\ie\pi I(E).
\ee
The density of states is symmetric in chiral and Bogoliubov-de Gennes classes
$\rho(-E)=\rho(E)$. Then besides $K(z^*)=K^*(z)$ also $K(-z)=K(z)+\ie\pi s(z)$
with $s(z)=\sign\Im(z)$ holds. If $K(z)+\ie\pi s/2\propto z^\gamma$ for small
$z$, then
\be
K(z)+\ie\frac{\pi s}2=cr^{\gamma}\ex{\ie\gamma(\phi-s\pi/2)}, \quad
z=r\ex{\ie\phi}
\ee
with real $c$. Then
\be
K(E+\ie 0) = c|E|^{\gamma} \left(\cos(\frac{\pi}2\gamma)
-\ie\,\sign(E)\sin(\frac{\pi}2\gamma)\right) - \frac{\pi}2\ie, \label{Kgamma}
\ee
from which $\lambda\propto|E|^{\gamma}$, $\rho(E)\propto|E|^{\gamma-1}$ follows
in agreement with (\ref{scalinglaw}).
One observes that
\be
\frac{\de K(z)}{\de z} = \int\de x \frac{\rho(x)}{z-x}. \label{diffK}
\ee
Thus the sign of the imaginary part of (\ref{diffK}) is opposite to the sign of
$\Im z$. This implies that $\gamma\le 1$. If a contribution with $\gamma>1$
appears, then there is also a contribution with $\gamma=1$, which according to
(\ref{Kgamma}) does not contribute to $\lambda$, but to a finite density of
states in the center of the spectrum. Such a system is described by the
homogeneous fixed point ensemble.

For $\gamma\le 0$ the integrated density of states would diverge. Thus these
arguments can only be applied for $0<\gamma<1$.

In a number of cases the asymptotic behaviour is given by a power multiplied by
some power of the logarithm. Then
\be
K(z) +\ie\frac{\pi s}2 = cr^{\gamma}\ex{\ie\gamma(\phi-s\pi/2)}
\left(\ln r+\ie(\phi-s\pi/2\right))^g.
\ee
This yields for $\gamma=0$ and $\gamma=1$
\be
\begin{array}{lcc}
\gamma && K(E+\ie 0) \\ \hline
0 && c(\ln|E|)^g -\ie cg\frac{\pi}2 \sign(E) (\ln|E|)^{g-1} \\
1 && -\ie cE(\ln|E|)^g -cg\frac{\pi}2|E|(\ln|E|)^{g-1}
\end{array}
\ee
and thus
\be
\begin{array}{ccccc}
\gamma && \lambda\sim && \rho\sim \\ \hline
0 && (\ln|E|)^g && \frac{(\ln|E|)^{g-2}}{|E|} \\
1 && |E|(\ln|E|)^{g-1} && (\ln|E|)^g
\end{array} \label{logoned}
\ee

Dyson\cite{Dyson53} calculated the averaged density of states for the chain
(\ref{Hchain}) with $\epsilon_i=0$ and random independently distributed matrix
elements $V$, for which he assumed a certain distribution and obtained
\be
\rho(E) \sim \frac 1{|E(\ln|E|)^3|}, \label{Dysonlaw}
\ee
which corresponds to the case (\ref{logoned}) with $\gamma=0$ and $g=-1$.
Indeed Theodorou and Cohen\cite{Theodorou76} and Eggarter and
Ridinger\cite{Eggarter} found the averaged localization length diverging
\be
\xi \sim |\ln|E||
\ee
in agreement with (\ref{logoned}).

\subsection{Ziman' s model}

Ziman\cite{Ziman} (compare also Alexander et al.\cite{Alexander}) considered a
one-dimensional tight-binding model (his case II)
(\ref{Hchain}) requiring the diagonal matrix elements to vanish,
$\epsilon_i=0$, and the hopping matrix-elements to agree pairwise
$V_{2m,2m+1}=V_{2m+1,2m+2}$. Apart from this restriction he assumed the $V$s to
be independently distributed with probability distribution
\be
p(V) = (1-\alpha) V^{-\alpha}, \quad 0<V<1, \quad -\infty < \alpha < 1.
\ee
He obtained for these distributions
\be
\begin{array}{ccccc}
 && \nu && \mu \\ \hline
-1<\alpha<1 && \frac{2(1-\alpha)}{3-\alpha} && \frac{-1-\alpha}{3-\alpha} \\
-3<\alpha<-1 && \frac{1-\alpha}2 && 0 \\
\alpha<-3 && 2 && 0
\end{array}
\ee
Obviously the first row describes models in accordance with the IHFPE, the
second and third row with the HFPE.

\subsection{Further one-dimensional results}

Titov et al.\cite{Titov} have summarized and completed results for the density
of states of all classes of chains with $N$ channels as shown in table
\ref{tb2}.

\begin{table}[h]
\tbl{Density of states close to $E=0$ for various universality classes
(after Titov et al.\cite{Titov})}
{\begin{tabular}{lll} \toprule
Class & & $\rho(E)\quad x=E\tau$ \\ \colrule
Chiral & all classes, odd $N$ & $|x^{-1}\ln^{-3}(x)|$\\
 & AIII, even $N$ & $|x\ln x|$\\
 & CII, even $N$ & $|x^3\ln x|$\\
 & BDI, even $N$ & $|\ln x|$\\ \colrule
BdG & CI & $|x|$\\
 & C & $x^2$ \\
 & D,DIII, $N\not=2$ & $|x^{-1}\ln^{-3}(x)|$ \\
 & D,DIII, $N=2$ & two mean free paths \\
\botrule
\end{tabular}}
\label{tb2}
\end{table}

All chiral classes are equivalent by a gauge transformation for $N=1$ and yield
the Dyson result (\ref{Dysonlaw}) and
$\xi\propto|\ln|E||$ for this case in agreement with (\ref{logoned}). 
Due to Gruzberg et al.\cite{Gruzberg05} also the BdG classes BD and DIII fall
into the same universality class.
The localization length does not diverge for the chiral classes, if $N$ is even.
The same holds for (general $N$) for the BdG classes C and CI.

\section{Bosons From One To Two Dimensions}

\subsection{One-dimensional chain}

Whereas the Hamiltonian (\ref{Hchain}) yields the equation for eigenstates
$|\psi\rangle = \sum_i \psi_i |i\rangle$
\be
E\psi_i = \epsilon_i \psi_i - V_{i,i-1} \psi_{i-1} - V_{i,i+1} \psi_{i+1}
\ee
one obtains a similar equation for harmonic phonons governed by the Hamiltonian
\be
H=\sum_i \frac{p_i^2}{2m} + \sum_i \frac{W_i}2 (x_{i+1}-x_i)^2,
\ee
which reads
\be
m\omega^2 x_i = W_i(x_i-x_{i+1}) + W_{i-1}(x_i-x_{i-1}).
\ee
Thus Thouless' arguments can be applied again with $x=\omega^2$. Since there are
no states for $\omega^2<0$, one has
\be
K(z) = cr^{\gamma} \ex{\ie\gamma(\phi-s\pi)}, \quad z=r\ex{\ie\phi},
\ee
which yields
\be
K(\omega^2+\ie 0) = c\omega^{2\gamma} \ex{-\ie\pi\gamma}
\ee
and thus
\be
\lambda(\omega) = c\omega^{2\gamma}\cos(\pi\gamma), \quad
I(\omega) = c\omega^{2\gamma}\sin(\pi\gamma), \quad
0<\gamma<1.
\ee
Alexander et al.\cite{Alexander} (cases a,b, and c) and Ziman\cite{Ziman} (case
II) determined the density of states $\rho(\omega)\propto\omega^{\mu}$ for
harmonically coupled
phonons with independently distributed spring constants
\be
p(W) = (1-\alpha) W^{-\alpha}.
\ee
Ziman moreover determined the localization length $\xi\propto\omega^{-\nu}$ and
obtained
\be
\begin{array}{c|cc}
 & \nu & \mu \\ \hline
 0<\alpha<1 & \frac{2(1-\alpha)}{2-\alpha} & -\frac{\alpha}{2-\alpha} \\
-1<\alpha<0 & 1-\alpha & 0 \\
\alpha<-1 & 2 & 0
\end{array}
\ee
Again the first line is in agreement with the IHFPE, whereas the two other cases
correspond to the HFPE.

\subsection{Bosonic excitations discussed by Gurarie and Chalker}

Gurarie and Chalker\cite{Gurarie03} point out that bosonic systems with and
without Goldstone modes show different localization behavior.

John et al.\cite{John83} investigated localization in an elastic medium with
randomly varying masses. For $d>2$ they found extended states for small
frequencies $\omega$. The phonon states are localized beyond some critical
$\omega_{\rm c}$. This transition is described by the orthogonal ensemble. For
$d<2$ all states are localized and obey\cite{John83} $\nu=2/(2-d)$. The density
of states for phonons shows the same power law
$\rho(\omega)\propto\omega^{d-1}$ as in the ordered case. In this system with
Goldstone modes the critical density below which the density of states would
differ from that of the ordered system is $d_{\rm c}=0$.

In a disordered antiferromagnet one obtains below the critical dimension $d_{\rm
c}=2$
\be
\rho(\omega) \sim \omega^{\mu}, \quad \mu=\frac{3d-4}{4-d}, \quad
\xi(\omega) \sim \omega^{-\nu}, \quad \nu=\frac 2{4-d},
\ee
where the result of \cite{Stinchcombe88} and the argument of\cite{Gurarie03}
have been generalized from $d=1$ to general $d<d_{\rm c}$. This is in agreement
with the IHFPE. These results rest on the assumption that there is a single
length scale $\xi\propto 1/k$.

\section{Electronic Systems In Two Dimensions}

\subsection{Conductivity in two dimensions\label{cond}}

From the homogeneity law $s=(d-2)\nu$, (\ref{scaling_s}), which works well for
$d>2$, I concluded\cite{Wegner76} $s=0$ for dimensionality $d=2$ and thus a
jump to a minimum metallic conductivity. At that time I did not expect that
$\nu$ may diverge as $d$ approaches 2. This was found three years later by means
of explicit renormalization group calculations\cite{Abrahams,Oppermann}.
Thus the idea of a minimum metallic conductivity was in error
for the orthogonal and unitary Wigner-Dyson classes, where all states are
localized in $d=2$. The critical conductivity in the symplectic class shows
some distribution\cite{Shapiro87} and is of order $e^2/h$.

In many two-dimensional models of chiral and Bogolubov-de Gennes
classes including the classes applying to d-wave superconductors and graphene
one obtains a finite non-zero conductivity of order $e^2/h$ at criticality.
This is to a large extend due to edge currents, as first observed by Pruisken et
al.\cite{Levine,Pruisken} for the integer quantum Hall effect. Thus although the
prediction\cite{Wegner76} turns out to be correct, the true mechanism is more
complex.

\subsection{Chiral and Bogolubov-de Gennes models in $d=2$ dimensions}

The unitary case of chiral models (Gade and Wegner, Gade\cite{Gade91,Gade93})
yields at intermediate energies effective exponents
\be
\nu=\frac 1B, \quad \mu=-1+\frac 2B
\ee
in accordance with (\ref{scalinglaw}). At asymptotically low energies
$\rho\propto E^{-1}\xi^2(E)$
corresponds to the limit $B\rightarrow\infty$. These results
as well as many similar results for various disordered Dirac hamiltonians are
obtained under the assumption that the localization length is given by the
cross-over length from chiral to Wigner-Dyson behaviour without taking further
renormalization into account\cite{Gade93}; see also the argument after eq.
(6.60) of the review by Evers et Mirlin\cite{Evers}. Alternatively the
integrated density of states from the band center up to energy $E$ is set to
$\xi^{-2}$ as in Motrunich et al.\cite{Motrunich02}, which yields
(\ref{scalinglaw}) per definition.
It is important that only one coupling yields a relevant
perturbation. The conductivity itself stays constant for the chiral models in
$d=2$. The exponent which drives the renormalization of the
energy is usually called dynamical exponent $z$, which is identical to the
exponent $y$ of (\ref{scalinglaw}). A more rigorous investigation of the
localization of such systems taking into account any dependence of the initial
couplings on the energy and of the cross-over would be of interest.

The spin quantum Hall effect yields at the percolation transition
point\cite{Saleur87,Gruzberg99,Beamond02}
\be
\nu=4/7, \quad \mu=1/7
\ee
in agreement with (\ref{scalinglaw}). The same behavior is obtained for the
Bogolubov-de Gennes class C if two of the four nodes of a dirty d-wave
superconductor are coupled\cite{Nersesyan,Altland02}.

\subsection{Power law for density of states, finite localization length}

The two fixed point ensembles describe the situation, in which the localization
length diverges and the density of states either approaches some finite
non-zero value (HFPE) or diverges or goes to zero by a power law, which may be
augmented by a logarithmic term. As mentioned above this holds for chains with
an even number of channels in the chiral classes and for the Bogolubov-de
Gennes classes C and CI. Certain single-channel models of class D and DIII show
a divergence of the density of states $\rho\propto|E|^{-1+\delta}$ without
divergence of the localization length\cite{Motrunich01}.

Gurarie and Chalker\cite{Gurarie02} found that bosonic excitations in random
media, which are
not Goldstone modes, obey $\rho\propto\omega^4$ with finite
localization length at low frequencies.

Apparently this type of behavior is not covered by HFPE and IHFPE.

\section{Conclusion}

The scaling prediction (\ref{scalinglaw}) of the IHFPE relating the exponent of the density of states and of the localization length yields correct results in the cases, in which I found both exponents.
The author appreciates the wealth of systems, which has been found and investigated over the years.

\section*{Acknowledgments}

The author enjoyed part of the summer program {\it Mathematics
and Physics of Anderson Localization: 50 years after} at the Newton Institute
of Mathematical Sciences in Cambridge. He gratefully acknowledges a Microsoft
Fellowship. He thanks for many useful discussions in particular with John
Chalker, Alexander Mirlin, Tom Spencer, and Martin Zirnbauer.

\end{document}